\documentclass{emulateapj}
\usepackage{apjfonts}

\makeatletter

\newcommand{\ros}{{\sl ROSAT}}
\newcommand{\chan}{{\sl Chandra}}

\newcommand{\xmm}{{\sl XMM-Newton}}

\newcommand{\fdeg}{\hbox{$.\!\!^{\circ}$}}

\def\aciss3{{ACIS-S3}}
\def\asca{{\sl ASCA}}

\def\fluxu{ergs cm$^{-2}$ s$^{-1}$}
\def\lumu{ergs s$^{-1}$}

\def\ls{\lower 2pt \hbox{$\;\scriptscriptstyle \buildrel<\over\sim\;$}}

\shorttitle{{\sl Chandra}  Observation of PSR B1823--13 }
\shortauthors{Pavlov, Kargaltsev, \& Brisken}

\makeatother
\begin{document}

\title{Chandra Observation of
PSR B1823--13
 and its Pulsar Wind Nebula}

\author{G.\ G.\ Pavlov\altaffilmark{1}, O.\ Kargaltsev\altaffilmark{1},
and W.\ F.\ Brisken\altaffilmark{2}}

\altaffiltext{1}{Dept. of Astronomy and Astrophysics,
Pennsylvania State
University, 525 Davey Lab., University Park,
PA 16802; pavlov@astro.psu.edu, oyk1@psu.edu}
\altaffiltext{2}{National Radio Astronomy Observatory, P.O. Box O, Socorro, NM 87801; wbrisken@nrao.edu}


\begin{abstract}
We
report on an observation of the Vela-like pulsar B1823$-$13 and
its synchrotron nebula with the {\sl Chandra X-ray Observatory}.
The pulsar's
spectrum fits a power-law model with a photon index $\Gamma_{\rm PSR}
\approx 2.4$ for the plausible hydrogen column density $n_{\rm H} =
1\times 10^{22}$ cm$^{-2}$, corresponding to the luminosity $L_{\rm
PSR}\approx  8\times 10^{31}$ ergs s$^{-1}$ in the 0.5--8 keV band, at a
distance of 4 kpc. The pulsar
radiation likely includes magnetospheric and thermal
components, but they cannot be reliably separated because of the small
number of counts detected and strong interstellar absorption.
The pulsar is
surrounded by a compact, $\sim 25''\times 10''$,
 pulsar wind nebula (PWN) elongated in the east-west direction, which
includes a brighter inner component
$\approx 7''\times 3''$,
elongated in the northeast-southwest direction.
The slope of the compact PWN spectrum is $\Gamma_{\rm comp}\approx 1.3$,
and the 0.5--8 keV luminosity is $L_{\rm comp}
\sim 3\times 10^{32}$
ergs s$^{-1}$.
The compact PWN is surrounded by asymmetric
diffuse emission (extended PWN) seen
up to at least $2.4'$ south of the pulsar,
with a softer spectrum
($\Gamma_{\rm ext} \approx 1.9$ for
$n_{\rm H}=1\times10^{22}$ cm$^{-2}$),
and the 0.5--8 keV luminosity
$L_{\rm ext}\sim 10^{33}$--$10^{34}$ ergs s$^{-1}$.
We also measured the pulsar's proper
motion using archival VLA data: $\mu_\alpha=23.0\pm 2.5$ mas yr$^{-1}$,
$\mu_{\delta}=-3.9\pm 3.3$ mas yr$^{-1}$, which corresponds to
the transverse velocity $v_\perp
\approx 440$ km s$^{-1}$.
The direction of the proper motion is approximately parallel to
the elongation of the compact PWN, but
it is nearly perpendicular to
that of the extended PWN and to the direction towards the center
of the bright VHE $\gamma$-ray source HESS J1825--137,
which is likely powered by PSR B1823$-$13.
\end{abstract}


\section{Introduction}

About a dozen of
$\sim 80$ radio pulsars observed in X-rays belong to the group of the so-called
Vela-like pulsars, with spindown ages $\tau\equiv P/2\dot{P} = 10$--30 kyr
and spindown powers $\dot{E}$ of a few $\times 10^{36}$ ergs s$^{-1}$.
Their radiation usually show
a thermal component,
 possibly emitted from the cooling neutron star (NS) surface,
and a nonthermal component with a power-law (PL) X-ray spectrum,
presumably originating in the pulsar magnetosphere (see Kaspi et al.\ 2006
for a review).
These pulsars are enveloped by pulsar wind nebulae
(PWNe) whose X-ray emission is due to the wind of ultrarelativistic
particles shocked in the ambient medium (Kargaltsev et al.\ 2007, and
references therein). Therefore, in addition to studying pulsar
magnetospheres and NS
cooling, X-ray observations of Vela-like pulsars
allow one to study pulsar winds, the ambient medium, and their interaction.

PSR B1823--13 (also known as PSR J1826--1334,
hereafter referred to as B1823) is a typical Vela-like
pulsar.
It was detected in the
Jodrell Bank radio survey (Clifton et al.\ 1992).
Its period, $P=104$ ms, and period derivative, $\dot{P}=7.5\times 10^{-14}$
s s$^{-1}$, imply the spindown age $\tau = 21$ kyr,
spindown power $\dot{E}=2.8\times
10^{36}$ ergs s$^{-1}$, and surface magnetic field
$B=2.8\times 10^{12}$ G.
The distance to the pulsar,
$d\approx 4$ kpc, was estimated from the
pulsar's dispersion measure (${\rm DM} = 231$ pc cm$^{-3}$)
using the models of Galactic electron distribution (Taylor \& Cordes 1993;
Cordes \& Lazio 2002).
Although such a young pulsar is expected to reside in a supernova remnant
(SNR), observations of this field with the Very Large Array (VLA)
have not found an SNR, nor a radio PWN
(Braun, Goss, \& Lyne 1989;
Frail \& Scharringhausen 1997;
Gaensler et al.\ 2000).

This pulsar is particularly interesting because it possibly powers
an extended ($\sim 1^\circ$, i.e. $\sim 70$ pc, in size)
Very High Energy (VHE) $\gamma$-ray source HESS J1825--137
(Aharonian et al.\ 2005,2006).
The $\gamma$-ray emission extends asymmetrically
south-southwest of B1823, with
the peak surface brightness
$\approx 10'$ from the pulsar.
Aharonian et al.\ (2006, hereafter A06) found a softening
of the
0.25--35 TeV spectrum with increasing distance from the pulsar
(hence an energy-dependent morphology), with a photon index
$\Gamma_\gamma$ increasing from $\approx 1.9$ to $\approx 2.5$.
These authors discuss a scenario in which the $\gamma$-rays are produced
by Compton upscattering of
the cosmic microwave background (CMB) photons off
the ultrarelativistic electrons supplied by the pulsar.
As the $\gamma$-ray luminosity of the source, $L_\gamma \sim 3\times 10^{35}\,
d_4^2\,\,\,
{\rm ergs\,\, s}^{-1}$ above 200 GeV [$d_4\equiv d/(4\,{\rm kpc})$],
is only a factor of 10 lower than the
pulsar's spindown power,
A06 propose that
the spindown power was significantly higher in the past,
when the electrons responsible for the currently observed
 TeV emission were produced
by the pulsar.

Although the field of B1823 has been
observed with \ros, \asca, and \xmm,
the X-ray spectrum of the pulsar and the structure of its surroundings
remained unclear because of the relatively low angular resolution of
those missions.
Finley
et al.\ (1996, F96 hereafter) observed B1823
with the \ros\/ observatory and,
despite the small number of source counts detected
(e.g., 127 counts from a $5'$ radius circle around the pulsar position
in the 14.7 ks PSPC exposure)
found evidence of a point source surrounded by a compact PWN
of $\sim 20''$ radius,
and a  dimmer
component
extending up to $\sim4'$ south-southwest of the pulsar.
Because of the
poor statistics, it was impossible
to measure the spectra.
For an upper limit on the hydrogen column
density, $n_{\rm H,22}\equiv n_{\rm H}/(10^{22}\,{\rm cm}^{-2}) < 4$,
F96 estimated the upper limit
fluxes at a few $\times 10^{-12}$ \fluxu\ level for the emission
components, corresponding to luminosities
$\lesssim (4$--$8)\times 10^{33} d_4^2$ erg s$^{-1}$ in the 0.5--2.4 keV range.
Sakurai et al.\ (2001) were able to detect the extended structure
around the B1823 position with {\sl ASCA},
but that observation lacked the spatial resolution to separate the components.

Gaensler et al.\ (2003, G03 hereafter)
observed B1823 with the \xmm\ EPIC detectors.  The \xmm\ resolution
turned out to be too low to
resolve the pulsar,
but G03 detected two components of
the PWN
(designated as G18.0--0.7 based on its Galactic coordinates)
associated with this pulsar: a ``core'' of a $30''$ extent, elongated
in the east-west direction, and
a large-scale diffuse component of lower surface brightness,
 $\sim 5'$ in extent,
seen mostly
to the south of the pulsar.
Fitting the core and diffuse component spectra with a PL model, G03 found
the photon indices $\Gamma_{\rm core} \approx 1.6$ and
$\Gamma_{\rm diff}\approx 2.3$, and 0.5--10 keV luminosities
$L_{\rm core}\approx 9\times10^{32}d_4^2$ \lumu,
$L_{\rm diff} \approx 3\times10^{33} d_4^2$ \lumu,
for a foreground hydrogen column density $n_{\rm H,22} = 1.2$.
They
suggested that the core
represents the pulsar
wind termination shock at a distance $R_s\approx 15''$ from the
the pulsar  while the diffuse component indicates the
shocked downstream wind. G03 propose that the asymmetric morphology of
the diffuse component with respect to the pulsar is the result of a reverse
shock from an associated SNR, which has compressed and distorted the PWN.
 They place limits on the pulsar's
thermal emission:
$kT < 0.147$ keV and
bolometric luminosity
$L_{\rm bol} < 8.7\times10^{33} d_4^2$ \lumu\
(as seen by a distant observer),
assuming that the NS emits as a blackbody of a radius $R=12$ km,
for
$n_{\rm H,22} < 2$.

Because of its superior
angular resolution,
 \chan\
is suited much better than any other X-ray observatory
 for studying small-scale
extended structures
(such as the core of the B1823 PWN)
and resolving
point sources
possibly projected onto such structures.
Therefore,
our group proposed a {\sl Chandra} observation of
B1823,
with the goal to separate the pulsar from the PWN,
measure their spectra, investigate the PWN morphology with the best
possible resolution, and compare the pulsar and PWN properties with
those of other Vela-like pulsars.
The data from that observation immediately showed that the pulsar
can indeed be separated from the PWN,
the ``core'' PWN is highly nonuniform in surface brightness,
and the overall PWN morphology is more
complicated than it had been presumed based on the previous low-resolution
X-ray observations.
These results
have been
reported at several conferences (Teter et al.\ 2003a,b; Gonzalez \& Kaspi
2006; Kargaltsev et al.\ 2006), but they have not been published in
a refereed journal.
In this paper, we decribe the analysis and results of
the \chan\ observations
and present infrared (IR) and radio images of the field (\S2).
As the interpretation of the PWN morphology strongly depends on the
direction and magnitude of the pulsar's proper motion, we also
report the proper motion measurement in \S3. Implications
of our findings for the pulsar and PWN physics, including the
VHE $\gamma$-radiation,
are discussed in \S4, while \S5 summarizes the results
of this work.

\section{Observations and Data Analysis}
\subsection{{\sl Chandra} ACIS observation}
 B1823 was observed with the Advanced CCD Imaging Spectrometer (ACIS)
aboard {\sl Chandra} on 2002 October 24
(ObsID 2830).
Although last 13 ks of the 40,836 s exposure suffered from flaring particle
background ($9.7\times 10^{-7}$ counts s$^{-1}$ per pixel, on average),
we have checked that filtering out the periods of elevated background
at the expense of shortening the exposure time would actually reduce
the signal-to-noise ratio for the pulsar and the compact PWN and only
very slightly increase it for the large-scale extended PWN;
therefore, we chose to use the whole exposure for our analysis.

 The
observation was carried out in Faint mode, and the target was
imaged on S3 chip, with the radio pulsar position
 $\approx35''$ from the optical axis, where the angular resolution
is still not degraded.
 The other ACIS chips
activated during this observation were S1, S2, S4, I2 , and I3.
The detector was operated in Full Frame mode, which provides time
resolution of 3.24 seconds.
We analyzed the data
reprocessed by the Chandra X-ray Center (CXC)
on 2006 October 11 (rev.\ 2; ASCDSVER 7.6.9).
The data were reduced using the Chandra Interactive Analysis of
Observations (CIAO) software (ver.\ 3.4; CALDB ver.\ 3.3.0.1).
We used the software package XSPEC (ver.\ 11.3.2) for the spectral
analysis.

\begin{figure*}[t]
 \centering
\includegraphics[width=7.2in,angle=0]{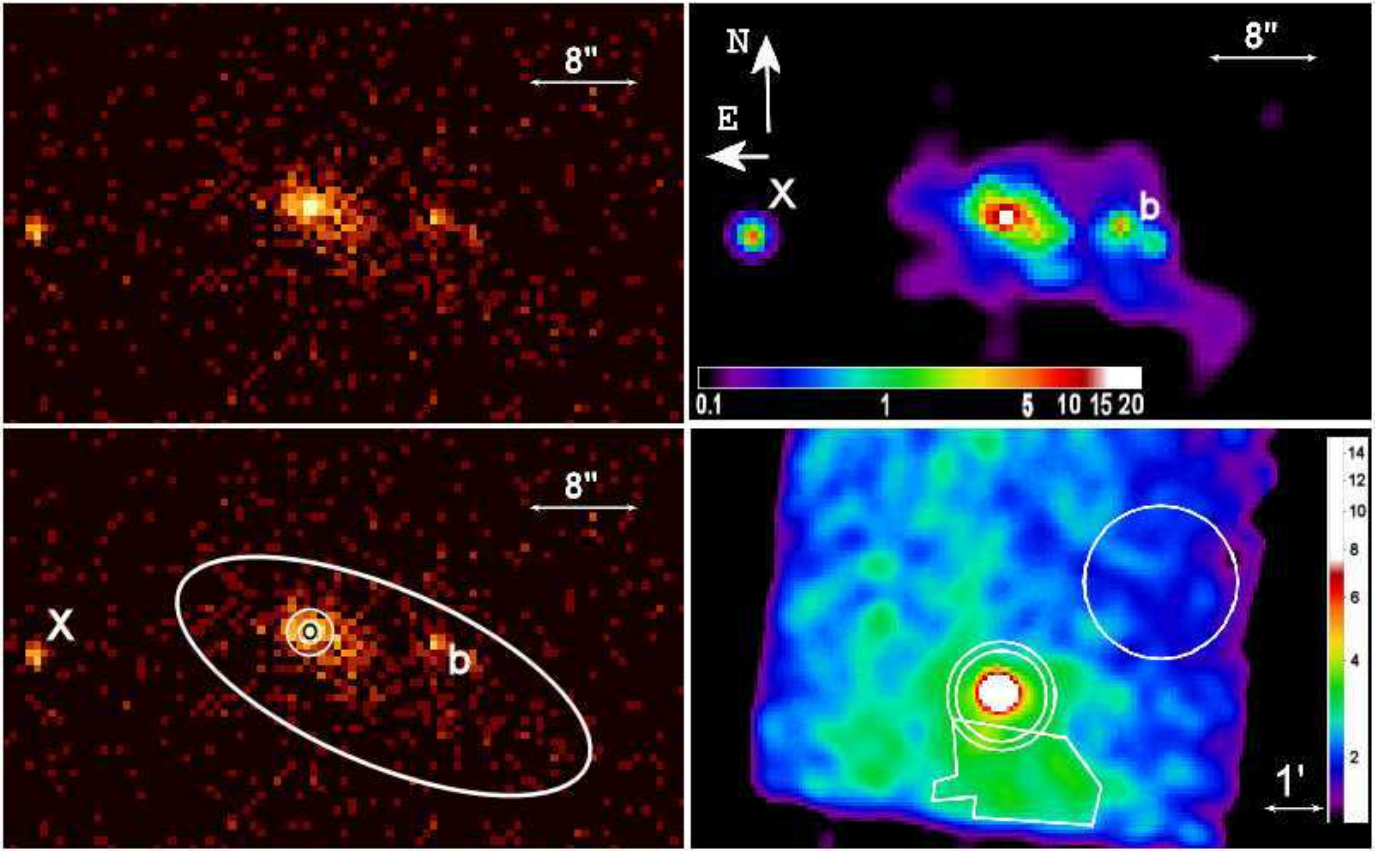}
\caption{ {\em Top left:} $77''\times 48''$ ACIS-S3 image
 of B1823 and its PWN (0.7--7 keV; pixel size $0\farcs49$).
{\em Top right:}
Adaptively smoothed
image (0.7--7 keV; pixel size $0\farcs49$)
 of the same region obtained by removing the
pipeline pixel-randomization and
applying the sub-pixel
resolution tool
(Tsunemi et al.\ 2001; Mori et al.\ 2001).
{\em Bottom left:}
 The same image as in the top left panel showing the extraction
regions used for the spectral analysis of the PWN and the pulsar (see
\S2). `b' and `X' mark the ``blob'' in the compact PWN and a point
source unrelated to the PWN, respectively.  {\em Bottom right:}
 Heavily binned (pixels size $3\farcs94$; $0.5-8.0$ keV) and
 smoothed (with a gaussian kernel of $r=27''$)
 ACIS-S3 image of B1823 and its surroundings.
The brightness and smoothing scales are chosen
 to stress the fainter, more extended emission south of the compact PWN.
The polygon shows the region from which we extract counts for the
spectral analysis of the large-scale extended emission, with the
background estimated from the $r=1.28'$ circle northwest of the
PWN. The annulus around the compact PWN is used for measuring
the compact PWN background.
}
\end{figure*}

\begin{figure}
 \centering
\includegraphics[width=3.2in,angle=0]{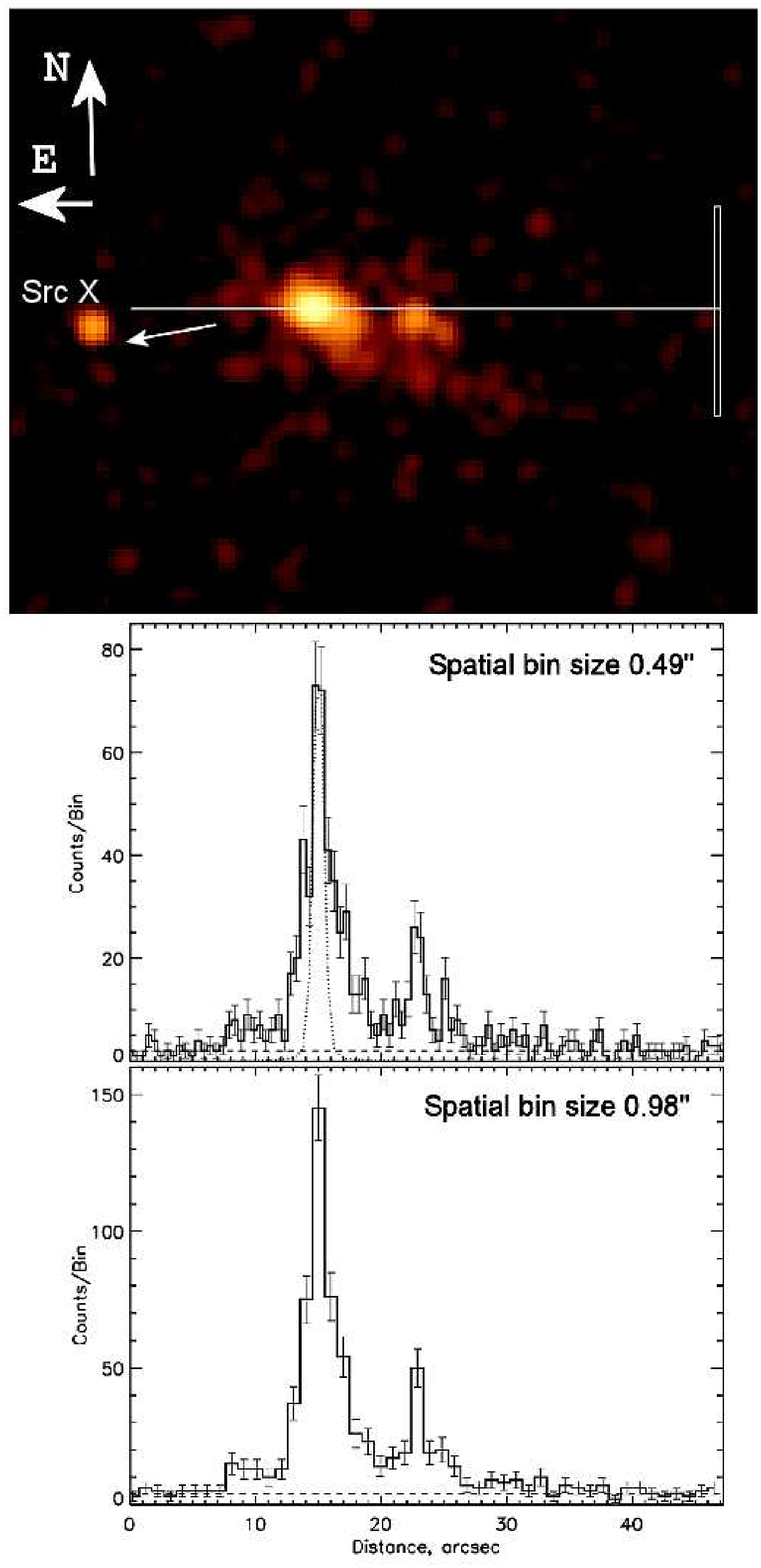}
\caption{
{\em Top:}
$62''\times  48''$
ACIS-S3 image ($0.7-7$ keV; pixel size $0.49''$; smoothed with $1.5''$
gaussian kernel)
of the immediate surroundings of B1823, including the compact PWN
and Source X  (likely a backlground AGN).
The
  white arrow shows the direction of the pulsar's proper motion
(see \S3).
The length of the arrow, $7''$,
  corresponds to the distance that the pulsar travels during
300 years.
 The {\em middle} and {\em bottom} panels show  the  linear profiles
extracted by moving the rectangular box
($0.49''\times17''$ and $0.98''\times17.1''$, respectively)
 along the horizontal line
  shown in the top panel. The one-dimensional point spread function
centered on the pulsar position is shown in the middle panel.
  }
\end{figure}

\subsection{X-ray images of B1823 and its vicinity }

Figure 1 shows the ACIS-S3 images of
the B1823 field, with
an extended source around the radio pulsar position.
The coordinates of the highest peak in the X-ray brightness distribution
(which is at the center of a $2\times 2$ pixel island, with
equal numbers of counts in each of the 4 pixels)
are
$\alpha=18^{\rm h}26^{\rm m}13.187^{\rm s}$,
$\delta =-13^{\circ}34' 46.63''$ (J2000),
with an estimated $1 \sigma$
centroiding error of $0.2''$ for each of the coordinates.
This position differs by $0.08''$ in right ascension and $0.17''$ in
declination from
the radio pulsar position,
$\alpha = 18^{\rm h}26^{\rm s}13.182^{\rm s}\pm 0.04''$,
$\delta =-13^{\circ}34'46.8'' \pm 0.2''$,
calculated from the original position in Hobbs et al.\ (2004)
for the epoch of the {\sl
Chandra} observation (MJD 52,571), using  the  proper motion
measurement described in \S3.
These differences are smaller than the combined errors
($0.3''$ in right ascension and $0.4''$ in declination, at the 68\% level)
of the
peak position measurement, radio position, and {\sl Chandra} absolute
astrometry.
Therefore, there are no doubts that the highest brightness peak corresponds
to the pulsar B1823 while the surrounding emission represents its PWN.

We see from Figures 1 and 2 that the PWN morphology is asymmetric
and nonuniform.
The brightest {\em inner} PWN component
(surface brightness in the range of 4--23 counts arcsec$^{-2}$ in 0.7--7 keV),
is $\sim 7''\times 3''$ (i.e.\ $0.14\times 0.06$ pc$^2$
at $d=4$ kpc) in size.
It is approximately centered on
the pulsar and
 elongated along the northeast-southwest direction
(position angle
${\rm P.A.}\sim 50^{\circ}$, measured east of north).

The inner component is
embedded in
a dimmer (average surface brightness $\approx 1.2$ counts
arcsec$^{-2}$ in 0.7--7 keV)
 {\em outer} component of the {\em compact PWN}, within a
$25''\times 10''$
area.
It is stretched along the east-west direction,
with the pulsar
being closer to its east end (i.e. the outer component is more
extended in the direction approximately opposite to the direction
of the pulsar's proper motion; see \S3).
Its surface brightness distribution is rather nonuniform,
with a prominent enhancement
$\approx8''$ west of the pulsar (we designate it as the ``blob''
and mark `b' in the figures).
 The smoothed subpixel-resolution image shown in Figure 1
({\em top right}) suggests that the
blob is connected to the
inner PWN component  by a faint ``bridge''.
We have checked that there are no optical-NIR counterparts to the blob in
the catalogs available (e.g.,
the nearest 2MASS\footnote{Two Micron All Sky Survey
(Skrutzkie et al.\ 2006).}
 counterpart is $6.5''$ away).

Another compact
source is seen in the images $\approx 18''$ east
of the pulsar. This object, CXOU 182614.4$-$133448
(hereafter referred to as Source X) is point-like. The analysis of
its spectrum (\S2.5) shows that it is
likely a background AGN,
unrelated to B1823.

The smoothed image in Figure 2 ({\em top})
 shows
marginal evidence of two $\sim 20''$-long ``tails'', apparently emanating
from the pulsar in the west-northwest and west-southwest directions.
The linear profiles in the middle and bottom panels of Figure 2
    show that the outer PWN structure is
discernible
above the local background level of
$\sim 0.2$ counts arcsec$^{-2}$
 up to
$\approx 7''$ and $\approx 23''$
east and west of the pulsar,
respectively.

 The compact PWN (comprised of the above-described inner and outer
components) is, in turn, immersed in an even
dimmer large-scale diffuse
   emission
apparent in heavily binned
images.
As it is natural to assume that this diffuse emission is also produced
by the wind from B1823,
we will call it the {\em extended} PWN.
It is mostly concentrated southward
   of the compact PWN, and it is discernible in the ACIS images
up to
the edge of the S3 chip (i.e.\ $\approx2.4'$ from the
   pulsar; Fig.\ 1, {\em bottom right}).
(It likely extends further into the S2 chip, but
the lower sensitivity of this chip
 does not allow a reliable analysis of the faint
diffuse
 emission.)
The extended
PWN is somewhat better seen in the {\sl XMM-Newton}
EPIC MOS1+MOS2 image shown in Figure 3{\em c}
(see also G03).
However,
the true extent of this
dim diffuse component is difficult to determine
even in the large field-of-view of the MOS detectors
because of the strong vignetting at large  off-axis angles and
   high EPIC background.

\begin{figure*}[t]
 \centering
\includegraphics[width=6.5in,angle=0]{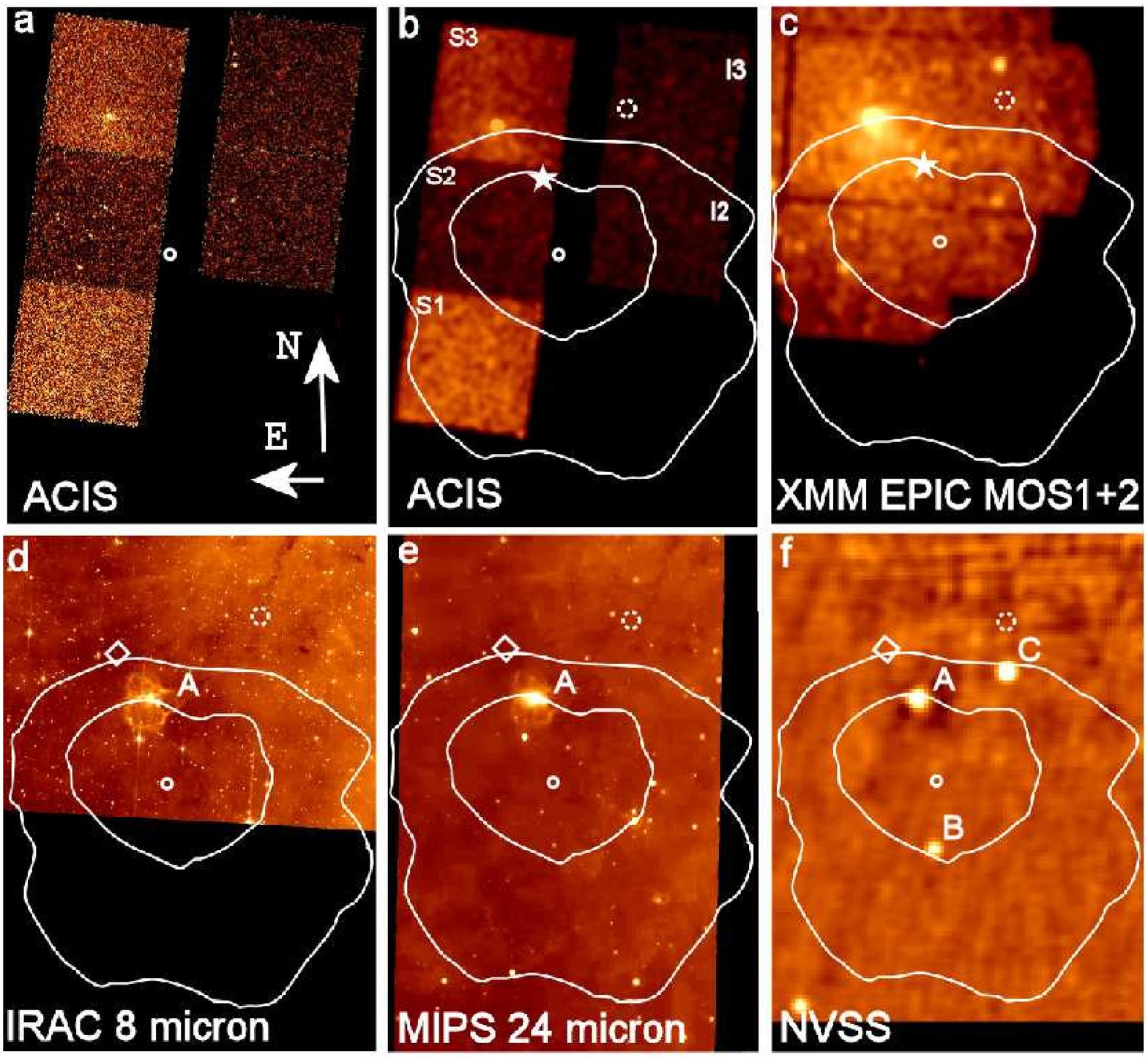}
 \caption{
Large-scale images of the B1823 vicinity and the central part
of HESS\,J1825
observed with different telescopes and at different wavelengths.
  All eight panels show the same area on the sky.   The TeV
brightness contours
 (adopted from
A06) are overlayed on top of
the images in panels {\em b--f}.
{\em a:} ACIS
S3, S2, S1, I2 and I3
image (0.7$-$7 keV; pixel size $3\farcs94$),
corrected for the exposure map nonuniformity.
{\em b:}
 The same image with point sources removed
(except for B1823)
and gaussian ($15.7''$ kernel) smoothing applied.
 {\em c:}   {\sl XMM-Newton} EPIC MOS1+MOS2  combined
image (0.5--8 keV; pixel size $10''$; smoothed with the $40''$ gaussian
 kernel).
{\em d:}
 {\sl Spitzer} IRAC 8\,$\mu$m image from the GLIMPSE survey.
{\em e:}  {\sl Spitzer} MIPS 24\,$\mu$m image from the GLIMPSE survey.
{\em f:} NRAO VLA Sky Survey (NVSS)
image at 1.4 GHz.
The star in panels {\em b} and {\em c} marks the location of
the brightest part of the extended
   IR and radio source A
 seen in the {\sl Spitzer} and NVSS images.
The diamond in {\em panels d--f} marks the B1823 position.
The small white circle
 marks the position of the brightness peak in the distribution
   of the extended TeV emission from HESS\,J1825.
 The projected place of birth of the pulsar estimated from the measured proper motion (see \S3)
 and pulsar's characteristic age ($\tau=21$ kyr) is shown by the dashed circle.}
\end{figure*}

\subsection{PWN spectrum}

We extracted the compact PWN spectrum\footnote{As we found no statistically
significant differences between the spectra
of the inner and outer components of the compact PWN, we present here
only the spectrum of the entire compact PWN.} (Fig.\ 4, {\em top})
from the
ellipse  shown in Figure 1 ({\em bottom left}),
excluding the counts
within the $r=3''$ circle centered on the X-ray position of the pulsar (see
\S2.1) and restricting the energy range to 0.7--7 keV.
This region of
225 arcsec$^2$ area
  contains
 415 counts.
We estimated the background from the annulus of
2916 arcsec$^2$ area
 shown in Fig.\ 1 ({\em bottom right}),
which contains
  511 counts. Thus, there are $375\pm 20$ compact PWN counts in the aperture
chosen.

To analyze the spectrum of the
extended PWN
(Fig.\ 4, {\em bottom}), we
chose the polygon encompassing the region of brightest
emission south of the pulsar,
with the area of  3.4 arcmin$^{2}$  (see Fig.\ 1, {\em bottom right}),
which contains
2329 counts in the 0.7--7 keV band.
The background was
extracted from the $r=1.28'$ circular region
shown in the same panel, which contains
 2004 counts. Subtracting the background contribution, we obtain
$1008\pm 57$ extended PWN counts in the polygon aperture.
(We have also tried several
other background regions but this
did not lead to significant changes in the
spectral fits.)

First, we fit the  compact and extended PWN spectra with the absorbed power-law
(PL) model, allowing the hydrogen column density
to vary. The confidence contours obtained from these fits are shown in
Figure 5.
We see that the 90\% confidence contours for the compact and extended PWN
do not overlap, which suggests that the two spectra are different.
Since we do not expect the hydrogen column density,
$n_{\rm H,22}\sim 1$, to differ significantly
on such a small, $\sim 2'$, scale,
we ascribe the difference to different spectral slopes,
$\Gamma_{\rm ext}- \Gamma_{\rm comp} \approx 0.5$.

To estimate the spectral slopes more accurately,
it would be useful to restrict
the range of $n_{\rm H}$
(which would also be helpful for the spectral analysis of the
pulsar -- see \S2.4). The fits with free $n_{\rm H}$ give
$n_{\rm H,22}
= 1.24\pm0.26$ and $n_{\rm H,22}=0.72\pm0.15$ for the compact
and extended PWNe, respectively (the uncertainties
correspond to the 68\%
confidence level for a single interesting parameter).
The  values
of $n_{\rm H,22}$
obtained by G03 using the {\sl XMM-Newton} data
are in the range
of 0.8--1.7, depending upon the
spectral model and extraction region chosen.
The estimate based on
the pulsar's dispersion measure (DM$=231$ cm$^{-3}$)
and the usual assumption of 10\% degree of ionization of
the interstellar medium
(ISM)
gives
$n_{\rm H,22}=0.7$.
This is close to the lower limit,
$n_{\rm H, 22}\gtrsim 0.7$, estimated from the observed colors of
the foreground ($d=2.9$ kpc) O6 star HD 169727,
projected $\approx 8'$ southeast
of the pulsar (F96).
Finally,
the total Galactic HI column in that direction is $n_{\rm HI,22}
=1.4$--1.5
(Dickey \& Lockman 1990).
Since the $n_{\rm H}$ values deduced from an X-ray spectrum
under the assumption of standard element abundances usually
exceed the $n_{\rm HI}$ measured from 21 cm observations by a
factor of 1.5--3 (e.g., Baumgartner \& Mushotzky 2005),
this $n_{\rm HI}$ value does not contradict to those estimated
from the X-ray fits and the dispersion measure.
 Thus,
with allowance for
different kinds of uncertainties, it seems reasonable to assume
$0.7 \lesssim n_{\rm H,22}\lesssim 1.3$,
with $n_{\rm H,22}\approx 1$ being the most plausible estimate.

 With the hydrogen column density fixed at $n_{\rm H,22}=1.0$,
the absorbed PL fits to the compact and extended PWN
yield photon indices
$\Gamma_{\rm comp}\simeq 1.3$ and $\Gamma_{\rm ext}\simeq 1.9$,
and 0.5--8 keV luminosities $L_{\rm comp}\simeq 3\times 10^{32}$
and $L_{\rm ext}\sim 9\times 10^{32}$ ergs s$^{-1}$
(see also Fig.\ 6 and
   Table 1; the extended PWN luminosity is for the polygon region).

\subsection{Pulsar spectrum}

Although the peak of the surface brightness distribution is
clearly dominated by the pulsar,
the surface brightness of the inner PWN
grows toward the pulsar's position, which introduces some ambiguity
in separating the pulsar contrubution from that of the PWN
and complicates the
measurement of the pulsar spectrum.
However, based on the surface brightness distribution near the peak
(see, e.g., the one-dimensional profiles in Fig.\ 2),
it seems reasonable to assume that the PWN
surface brightness reaches a plateau of about $3''$ in size,
centered on the pulsar, which can be used to estimate the
local background at the pulsar position.
Therefore, we extracted the pulsar
spectrum from a small circular aperture (small black circle in Fig.\ 1,
{\em bottom left}) with the radius of 1 ACIS pixel ($\simeq 0.49''$,
55\% encircled energy radius)
and took the background
from
the $0.74''<r<1.5''$ circular annulus
centered on the pulsar
(shown by the
white circles in Fig.\ 1, {\em bottom left}).
 The background
contributes $\approx 22$ counts
to 62 counts extracted from the source region.
 Given the
small number of source counts
 and the appreciable background contribution,
we chose not to subtract the background channel by channel but
rather fit the background spectrum (modeled as an absorbed PL model
with the same $n_{\rm H}$ as for the source) simultaneously with the
source spectrum.

To obtain better constrained fits with the small number of counts
available, we are forced to freeze the hydrogen column density.
Since the interpretation of the pulsar spectrum (e.g., thermal vs.\
nonthermal) may depend on the choice of the $n_{\rm H}$ value,
 we  fit the
pulsar spectrum with the absorbed PL model for three
plausible
values of the hydrogen column density:
 $n_{\rm H,22}=0.7$, 1.0, and 1.3 (see Table 2 and Fig.\ 7).
As expected, the inferred photon indices depend on the
choice of $n_{\rm H}$, changing from $\Gamma_{\rm
PSR}=2.0\pm0.4$ to $3.0\pm0.6$ with increasing
$n_{\rm H}$ (see
Table 2 for details).
 The pulsar's absorbed flux
is $F_{\rm PSR}=(1.6\pm0.4) \times10^{-14}$ ergs cm$^{-2}$ s$^{-1}$
in the 0.7--7 keV band (corrected for the background and finite
aperture).
 The unabsorbed pulsar luminosity in the 0.5--8
keV band increases from $L_{\rm PSR} \approx 0.6\times 10^{32}$ to $1.1
\times 10^{32}$ ergs cm$^{-2}$ s$^{-1}$, respectively.

The steepness of the best-fit PL spectra, especially in the case of
larger $n_{\rm H}$, indicates the presence of thermal emission from
the NS surface, such as seen in other Vela-like pulsars (e.g.,
Pavlov et al.\ 2001).
However, adding the
second blackbody (BB) component is not justified statistically (i.e.\ the
contribution of the BB component is negligible, and the fitting
parameters of the BB component are unconstrained) at $n_{\rm
H,22}=0.7$ and 1.0, while in the case of
 $n_{\rm H,22}=1.3$ the
fit
 converges
to a reasonable (although still poorly constrained)
 values of the surface temperature $kT\simeq 100$ eV,  projected
 emitting area $\mathcal{A}\sim 500d_4^2$ km$^{2}$
(corresponding to the bolometric luminosity
$L_{\rm bol} \equiv 4{\mathcal{A}}\sigma T^4
\sim 2\times 10^{33}
d_4^2$ ergs s$^{-1}$),
 and the slope of the PL component $\Gamma_{\rm PSR}
\simeq 1.9$.
The fitting parameters become better constrained if we fix the
area, but
the best-fit temperature and luminosity depend on the value of
$\mathcal{A}$
because of the strong $\mathcal{A}$-$T$ correlation.
For instance,
$kT$ increases from 97 to 139 eV, and $L_{\rm PSR}^{\rm bol}$
decreases from $1.8\times 10^{33}$ to $3.0\times 10^{32}$ ergs s$^{-1}$,
  when $\mathcal{A}$ decreases
from 500 km$^{2}$ (approximate projected area of the NS surface)
to 20 km$^{2}$ (projected area in the PL+BB fit for the Vela pulsar
spectrum).
Given the errors of these fits and the uncertainty of ISM
absorption,
 the obtained BB parameters
are just crude estimates.
We can, however, treat
$L_{\rm bol}\sim 2\times 10^{33}
d_4^2$ ergs s$^{-1}$
  as
an upper limit for the bolometric luminosity.

\begin{figure}
 \centering
\includegraphics[width=2.9in,angle=0]{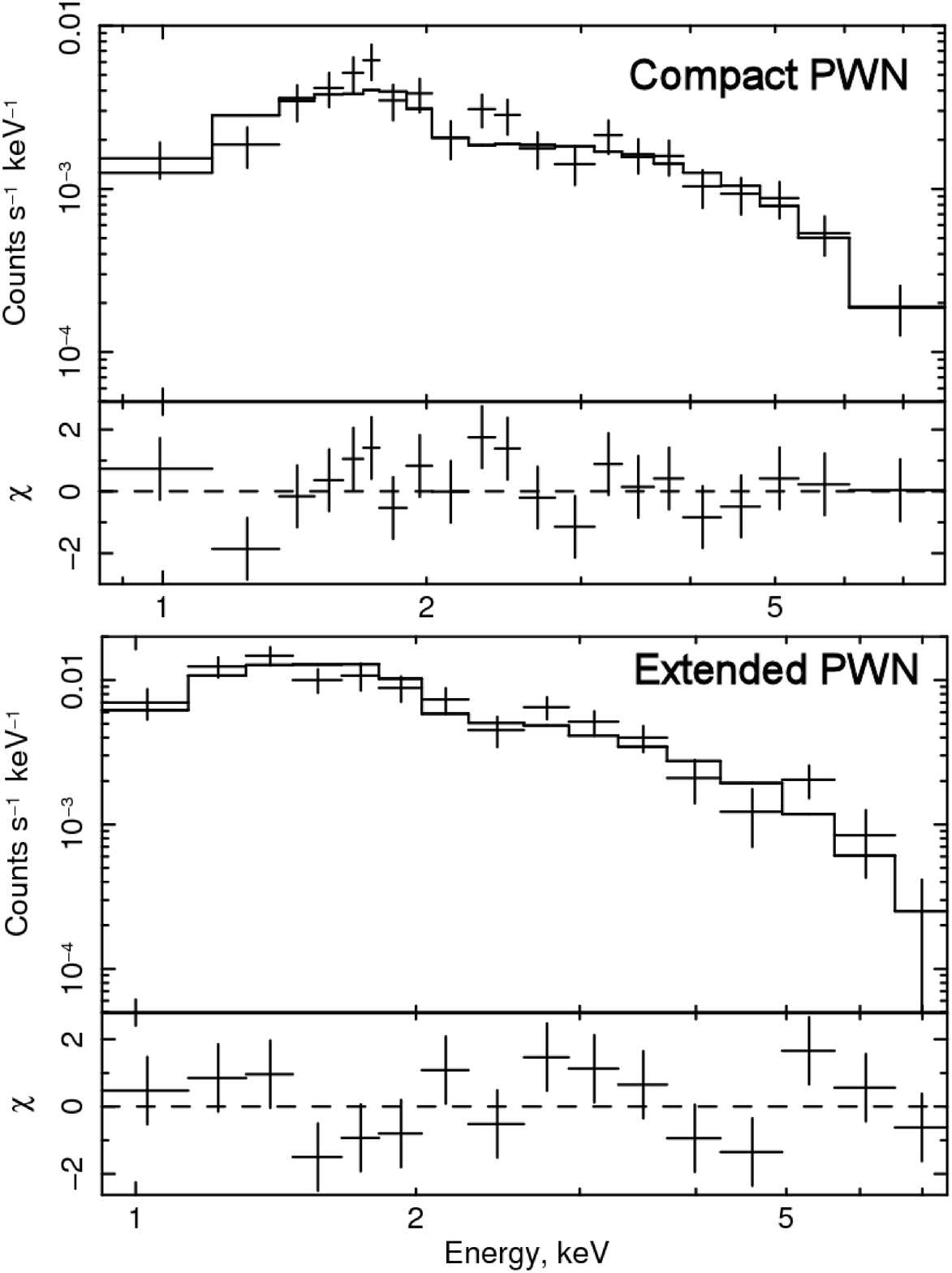}
 \caption{ Spectrum of the compact PWN ({\em top})
and the
extended PWN ({\em bottom})
 fitted with the PL model at fixed $n_{\rm H,22}=1.0$.
 }
\end{figure}

\begin{figure}
 \centering
\includegraphics[width=2.5in,angle=90]{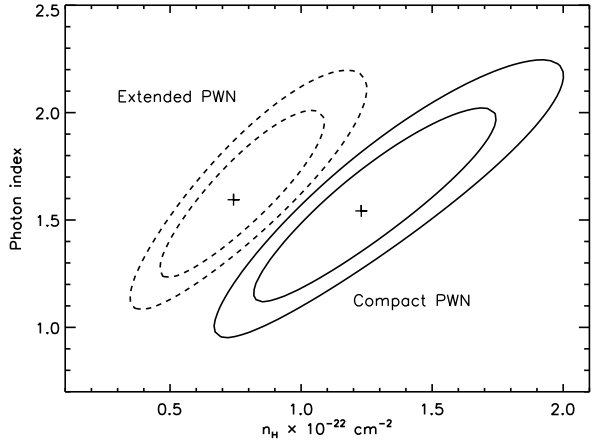}
\caption{Confidence contours (68\% and 90\%) in the $n_{\rm
H}$--$\Gamma$ plane for the PL fit to the
 compact and extended PWN spectra.
 The contours
 are obtained with the PL normalization fitted at each point
of the grid.
}
\end{figure}

\begin{figure}
 \centering
\includegraphics[width=2.5in,angle=90]{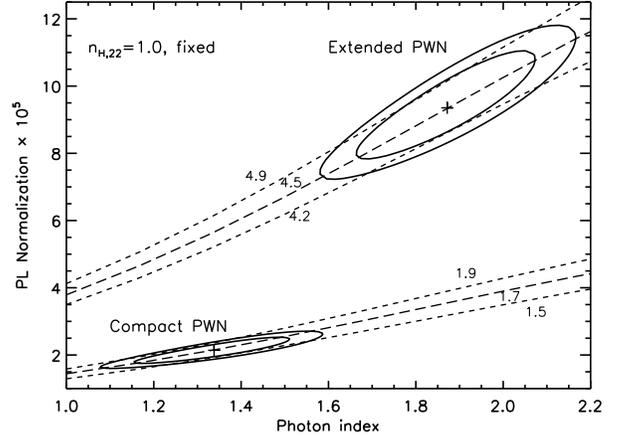}
\caption{  Confidence contours (68\% and 90\%) for the PL fit to the
 compact PWN and extended PWN
for fixed $n_{\rm H,22}=1.0$.
 The PL normalization
is in units of $10^{-5}$ photons cm$^{-2}$ s$^{-1}$ keV$^{-1}$ at 1
keV. The dashed curves are the loci
 of constant unabsorbed flux in the 0.5--8 keV band;
the flux values near the curves are in units of 10$^{-13}$ ergs
cm$^{-2}$ s$^{-1}$.
}
\end{figure}

\begin{figure}
  \centering
 \vbox{
\includegraphics[width=2.5in,angle=90]{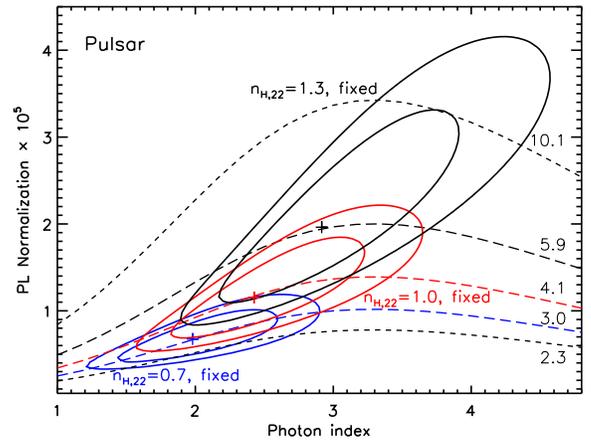}
} \caption{ Confidence contours (68\% and 90\%) for the PL fit to the pulsars's spectrum.
 The PL normalization
is in units of $10^{-5}$ photons cm$^{-2}$ s$^{-1}$ keV$^{-1}$ at 1
keV. The dashed curves are the lines of constant unabsorbed flux
in the 0.5--8 keV band (the flux values are in units of $10^{-14}$
ergs cm$^{-2}$ s$^{-1}$).
}
\end{figure}

\subsection{Other X-ray sources}

Among X-ray
sources serendipitously detected in this ACIS observation, most interestng are
CXOU\,182614.4$-$133448 (Source X, see Figs.\ 1 and 2) $\approx 18''$ east of
the B1823 pulsar,
and a few sources
 on the S2 and I2 chips projected within the brightest part of
HESS\,J1825 (see Fig.\ 8),
which might supply additional ultrarelativistic
electrons to power HESS\,J1825. The positions
and spectral characteristics of these sources
are presented in Table 3.

The strongly absorbed ($n_{\rm H,22}\sim 3$) point-like Source X
shows a hard spectrum ($\Gamma\sim 0.2$) and the
1--8 keV flux
of $\sim 3\times 10^{-14}$ ergs cm$^{-2}$ s$^{-1}$
(see Table 3 for other properties). There is no
an obvious optical-NIR counterpart for this source in stellar catalogs.
2MASS images show a point source with $J=13.4$, $H=12.3$ and
$K_s=11.85$ close to Source X
($\alpha=18^{\rm h}26^{\rm m}14.32^{\rm s}$,
$\delta=-13^\circ 34'48.2''$),
but, since the offset of $1.6''$ from the pulsar position exceeds the
combined centroiding and absolute astrometry error
($0.7''$ at the 90\% confidence level), the 2MASS source is unlikely
to be a counterpart of Source X.
Based on the X-ray spectrum, we conclude that Source X
is, most likely,
a background AGN, not related to either B1823 or HESS\,J1825.

The other X-ray sources listed in Table 3 also look point-like
(the apparent extension of the sources s4 and i1 can be
explained by the asymmetric PSF broadening at the large off-axis distances).
2MASS and Digital Sky Survey (DSS\footnote{see http://archive.eso.org/dss/dss})
 images do not show NIR-optical objects within most
conservative error circles around
the sources s1, s5, and i1 (the nearest objects are offset by
$4.1''$, $4.0''$, and $3.4''$, respectively).
The strongly absorbed s1 and s5 are likely extragalactic sources
(possibly AGNs),
while the nature of i1 is unclear. The source s4 shows a possible
 NIR-optical
counterpart DENIS\,J182620.9$-$134425
 (offsets are in the range of $0.36''$--$0.86''$ in different catalogs;
 magnitudes $K_s=11.75$, $H=12.2$, $J=13.3$
in the 2MASS catalog; $I=15.0$, $R=15.8$--16.4, $V=16.5$--17.1, $B=17.5$--19.1
in other catalogs). The source looks slightly extended in the NIR-optical
images, likely a blend of two field stars.
Close to the sources s2 and s3, there are
point-like
NIR-optical objects (DENIS J182617.1$-$134111
and J182618.0$-$134225, respectively),
but their relatively large
offsets from the X-ray positions
 ($1.4''$--$1.9''$ and $1.8''$--$2.3''$,
respectively, somewhat different in different catalogs)
make their association with the X-ray sources questionable.
Overall, the X-ray sources in the brightest part of HESS\,J1825
are rather faint and unremarkable,
and there is no evidence that any of the sources is related
to the VHE $\gamma$-ray source HESS\,J1825.

\begin{figure}[]
 \centering
\includegraphics[width=3.5in,angle=0]{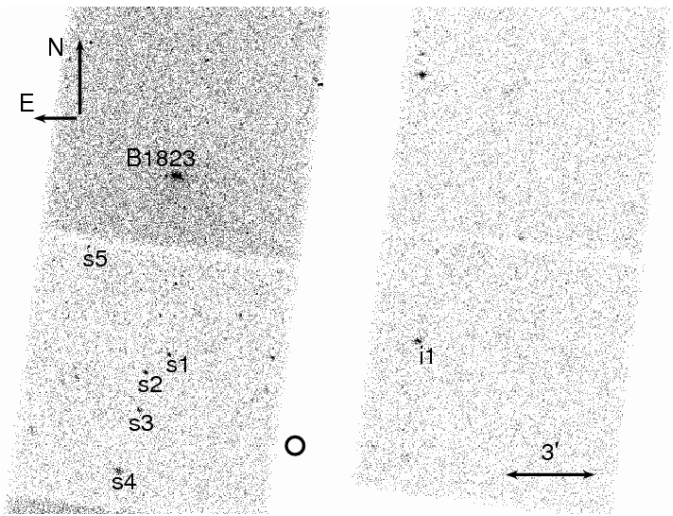}
\caption{ACIS field sources in the vicinity of the HESS\,J1825 center (shown by the circle).
}
\end{figure}

\subsection{Vicinity of PSR B1823-13 and HESS J1825--137 at other wavelengths}
To check whether or not there are luminous sources within the central part
of HESS\,J1825, which could enhance the VHE $\gamma$-radiation by providing
additional photons for the inverse Compton (IC) scattering
off ultrarelativistic electrons,
we have examined optical-NIR, IR, and radio images of this field.
Three examples, {\sl Spitzer} IRAC 8\,$\mu$m and MIPS 24\,$\mu$m
images from the Galactic Legacy Infrared Mid-Plane Survey
(GLIMPSE\footnote{http://www.astro.wisc.edu/sirtf/}) and a NRAO
VLA Sky Survey (NVSS; Condon et al.\ 1998) 1.4 GHz image are shown
in Figure 3. The images do not show anything remarkable in the
immediate vicinity, within $\sim 2'$, of B1823. The most interesting
object within the brightest part of HESS\,J1825 is the extended
($\approx 4'$ in size) source, marked A in Figure 3,
prominent in both the IR and radio images.
The source with a two-sided shell morphology is centered at
$\alpha \approx 18^{\rm h}26^{\rm m}02^{\rm s}$,
$\delta \approx -13^{\circ}38'08''$
(measured in the IRAC 8\,$\mu$m image), i.e.\
  between the pulsar and the brightest part of HESS\,J1825.
It is listed in the IRAS catalog of point sources\footnote{Infrared
Astronomical Satellite Catalog, 1988, The Point Source Catalog, ver.\ 2.0,
NASA RP-1190.} as IRAS\,18231--1340, with fluxes of
4.25, 13.4, 161, and 393 Jy at 12, 25, 60, and 100 $\mu$m, respectively.
Its radio fluxes are 0.13 and 0.03 Jy at 4.85 and 1.4 GHz, respectively
(Griffith et al.\ 1994; Condon et al.\ 1998).
This object is barely visible in the IRAC 4.5\,$\mu$m image, and it is
not seen in the
2MASS images of the field. The IR-radio spectrum and
   the lack of an extended optical or NIR counterpart
suggest that
the IR emission is thermal
emission of
dust with $T\sim 30$--50 K, possibly mixed with the
molecular/atomic cloud(s) that are heated and partly ionized by
radiation from embedded young stars.
Its luminosity can be crudely estimated as
$L_A\sim 3\times 10^{37}d_4^2$
ergs s$^{-1}$.

The nature of the other two bright radio sources within the brightest part
of HESS\,J1825 (B and C in Fig.\ 3{\em f}), centered at
$\alpha\approx 18^{\rm h}25^{\rm m}57^{\rm s}$,
$\delta\approx -13^\circ 47'56''$ and
$\alpha\approx 18^{\rm h}25^{\rm m}38^{\rm s}$,
$\delta\approx -13^\circ 36'37''$, respectively, is unknown.
However, as they are not seen in IR, their bolometric luminosity
is much lower than that of Source A.

\section{Proper motion of PSR B1823--13}

We measured the proper motion
using two unrelated archival
VLA observations.  Both observations were made in the 20~cm band in
the A-configuration providing about $1.5''\times 2.1''$ resolution.
The reduction technique
is identical to that used by
Brisken et al.\ (2006)
in the determination of the proper motion of PSR B1800--21 using
1.4 GHz data.
All synthesis calibration and imaging were
performed using the AIPS software package.
Project AF118 was observed on MJD 46573
in right circular polarization in spectral line mode providing
15 channels spanning 50~MHz bandwidth at 1385~MHz.
The upper 4 channels were discarded due to severe interference.
Project AC761 was observed on MJD 53265 in dual circular polarization
in spectral line mode providing 15 channels across 25~MHz at 1665~MHz.
After standard calibration and editing, AIPS task UVFIX was used to
correct the baseline vectors of the pulsar field for the half-integration time
time-stamp error
 common to all VLA observations as well as including
relativistic aberration correction.  UVFIX was also used to convert the
B1950 coordinates used in AF118 to the J2000.0 coordinates.

The temporal and spectral resolution of both data sets allowed imaging of
the entire 25\arcmin\ primary beam.  In addition to the pulsar, eleven other
sources were detected in both epochs, five of which were bright and
compact enough to be used as reference sources.  Accurate positions for
the pulsar and these five objects were measured for each epoch.
The motion of the pulsar was then measured in a frame defined by
these sources using a fitting prodedure described in
McGary et al.\ (2001), yielding a proper motion
$\mu_{\alpha} = 22.7 \pm 2.2$~mas~yr$^{-1}$,
$\mu_{\delta} = -5.3 \pm 2.7$~mas~yr$^{-1}$
(68\% confidence intervals). The global linear fit used to determine
the proper motion and a possible translation and rotation of the coorinates
between the two epochs yielded a reduced $\chi^2$ of 1.3,
indicating that the formal uncertainties may be underestimates at the
15\% level.  Assuming a distance of $4 \pm 1$~kpc,
 this proper motion implies a local standard of rest (LSR) proper
motion of $\mu_{\alpha} = 23.0 \pm 2.5$~mas~yr$^{-1}$,
$\mu_{\delta} = -3.9 \pm 3.3$~mas~yr$^{-1}$, which corresponds
to the velocity $v_\perp = (443\pm 48) d_4$ km s$^{-1}$ at a position angle
of $99\fdeg6\pm 8\fdeg1$ (east of north).
At this proper motion, the pulsar has traveled about $8.2'$ in the plane
of the sky during its characteristic age of 21 kyr. The projected
birthplace of the pulsar (hence the SN location) is marked in Figure 3{\em b-f}.

\section{Discussion}

\subsection{Pular Wind Nebula}
From their analysis of the {\sl XMM-Newton} data,
G03 have concluded that the B1823 PWN
(which they designated as G18.0--0.7)
is comprised
of two components (``core'' and ``diffuse'') of different brightness and
 size and perhaps different
spectra. Our {\sl Chandra} ACIS observation has confirmed the presence
of the two components, which we designate as the ``compact PWN''
and ``extended PWN'',
and allowed us to investigate the properties of the compact PWN in much more
detail.
The better resolved images and the measurement of
the pulsar's proper motion not only allow a more definitive
interpretation of the compact PWN properties,
but they also help better constrain the models
the puzzling extended PWN and its connection to HESS\, J1825.

\subsubsection{Compact PWN}

The east-west elongation of the compact PWN, approximately coinciding
with the direction of the pulsar's proper motion (see Fig.\ 2), strongly
suggests that the compact PWN
is shaped by the oncoming flow of the ambient matter in the pulsar's
reference frame.
The transverse velocity
of the pulsar, $v_\perp \approx 440 d_4$ km s$^{-1}$, is high enough to
assume that the pulsar's motion is supersonic,
i.e., ${\mathcal M}\equiv v/c_s >1$, or, equivalently,
$T< 1.4\times 10^7 \mu (d_4/\sin i)^{2}$ K, where
${\mathcal M}$ is the Mach number, $v=v_\perp/\sin i$ is the pulsar's
speed, $i$ is the angle between the velocity vector and
the line of sight, $c_s=(5kT/3\mu m_{\rm H})^{1/2}=
117 \mu^{-1/2} T_6^{1/2}$ km s$^{-1}$ is the speed
of sound, $\mu$ is the molecular weight,
and $T=10^6 T_6\,{\rm K}$ is the temperature of the ambient medium.
When a pulsar moves supersonically, the pulsar wind (PW) termination
shock (TS) acquires a bow-like shape ahead of the pulsar,
with the apex at a distance $R_h=(\dot{E} f_\Omega/4\pi c p_{\rm ram})^{1/2}$,
determined by the balance between the PW pressure and the
ram pressure, $p_{\rm ram}=\rho v^2$, where $\rho$ is the density of
the ambient medium, and $f_\Omega$ takes into account anisotropy of
the pulsar wind.
For B1823, these equations yield
$p_{\rm ram}=3.3\times 10^{-9} n (d_4/\sin i)^{2}$ ergs cm$^{-3}$ and
$R_h=4.8\times 10^{16} n^{-1/2} f_\Omega^{1/2} (d_4/\sin i)^{-1}$ cm,
which translates into the projected angular distance
${\mathcal R}_h \approx 0.8'' f_\Omega^{1/2} n^{-1/2}(d_4/\sin i)^{-2}$,
where $n=\rho/m_{\rm H}$.

In an idealized picture,
the shocked PW,
responsible for the PWN X-ray emission,
is confined between the TS and
the contact discontinuity (CD) surface, which separates the shocked
PW from the shocked ambient medium.
For ${\mathcal M}\gg 1$
and a nearly isotropic
preshock PW with a low magnetization parameter $\sigma$,
the distance between the pulsar and the head of the CD surface is
$\approx 1.3 R_h$ (Bucciantini et al.\ 2005, hereafter B05).
In a more realistic picture,
one should not expect a smooth CD surface because it is unstable.
As a result, the shocked PW can penetrate into the shocked ambient medium,
and vice versa.

The images of the compact PWN (Figs.\ 1 and 2)
do not show a sharp PWN boundary ahead of the pulsar that could be
unambiguosly identified as a CD surface head.
Instead, we see a two-component structure, with the bright inner
component up to $3''$ east of the pulsar and some fuzzy emission
discernible up to $7''$.
Although
the quality of the image does not
allow one to make a definitive conclusion, we can assume that the
fuzzy emission represents the shocked PW that has penetrated into the
shocked ambient medium because of CD instabilties, while
the inner component is emitted by the shocked PW immediately
beyond the TS, so that ${\mathcal R}_h\sim 2''$--$3''$ seems to be
a reasonable estimate for the angular distance to the TS head.
Scaling ${\mathcal R}_h$ to $3''$, we can estimate the ambient
number density,
$n\sim 0.07 f_\Omega (d_4/\sin i)^{-4}({\mathcal R}_h/3'')^{-2}$ cm$^{-3}$,
and the ram pressure,
$p_{\rm ram}\sim 2.3\times 10^{-10} f_\Omega (d_4/\sin i)^{-2}
({\mathcal R}_h/3'')^{-2}$ ergs cm$^{-3}$.
Such a ram pressure is typical for Vela-like pulsars (Kargaltsev
et al.\ 2007), while
such a number density can be found in a warm phase of the ISM or
within an evolved SNR. As no SNR emission is seen in the vicinity
of B1823, we cannot measure the temperature and pressure of the
ambient medium, but it seems reasonable to assume $T\sim 10^5$ K,
which could explain the lack of X-ray SNR emission. Scaling the
temperature to this value, we obtain $p_{\rm amb} = nkT/\mu \sim
1\times 10^{-12} f_\Omega (d_4/\sin i)^{-4}({\mathcal R}_h/3'')^{-2}\mu^{-1}
 T_5 \ll p_{\rm ram}$,
$c_s \sim 40 \mu^{-1/2} T_5^{1/2}$ km s$^{-1}$,
and ${\mathcal M} \sim 10 (d/\sin i) \mu^{1/2} T_5^{-1/2}$.

The existing MHD models of bow-shock PWNe associated with supersonically
moving pulsars assume a nearly isotropic preshock PW.
 However, the high-resolution {\sl Chandra} images of the whole
compact PWN can hardly be reconciled with this assumption. According
to B05, the TS in an isotropic PW of a supersonically moving pulsar
 acquires a bullet-like shape,
with the bullet's cylindrical radius $r_{\rm TS} \sim R_h$ and
the distance of its back surface from the pulsar $R_b\sim 6 R_h$.
The shocked PW outside the TS is confined inside the CD surface,
which has a cylindrical shape behind the TS bullet, with a
cylindrical radius $r_{\rm CD}
\sim 4 R_h$. The collimated PW flows with subrelativistic velocities:
$0.1c$--$0.2c$ in the inner channel, $r\lesssim r_{\rm TS}$, and
up to $0.8c$--$0.9c$ in the outer channel, $r_{\rm TS}\lesssim r \lesssim
r_{\rm CD}$ (see Figs.\ 1--3 in B05). The images of the compact PWN
are not quite consistent with this simple picture. First of all, the
bright inner component, which could be naturally associated with
the PW immediately outside the TS, is inclined to the direction
of the proper motion by an angle $\sim 50^\circ$, while it should be
coaligned with the proper motion if the wind is isotropic. Second,
the outer PWN, which might represent the shocked PW confined by
the CD surface, not only has an irregular boundary (which could be
explained by instabilities), but it is also rather nonuniform.
(Particularly puzzling is the blob $\approx 8''$ west of the pulsar;
see Figs.\ 1 and 2, and \S2.2). Finally, we do not see an extended
tail behind the TS aligned with the proper motion direction.
These differences between the model and the observed image suggest
that the preshock PW is, in fact, substantially anisotropic.
To understand the PW geometry and interpret the PWN morphology more
or less unambiguously, deeper high-resolution X-ray observations
are needed, as well as modeling of intrinsically anisotropic,
magnetized
PWs from fast-moving pulsars. Here we can only speculate that
the inner PWN likely represents a shocked equatorial outflow (similar to
the famous Crab torus; Weisskopf et al.\ 2000), seen almost
edge-on. If this is the case,
then the spin axis of the B1823 pulsar is strongly misaligned
(by $\sim 40^\circ$--$50^\circ$ in the plane of the sky)
with the direction of its motion, contrary to many other pulsars.
The absense of jets along the spin axis in the observed images
can be explained by their faintness (e.g., the jets of the Vela
PWN would be unobservable at $d\sim 4$ kpc with such an exposure --
see Fig.\ 8 in Kargaltsev et al.\ 2007), or possibly the jets
are deflected (or even crushed) by the oncoming flow of the ambient
medium. The blob and the apparent wiggles in the west part
of the outer component could be ascribed to instabilities
in the tail formed by the supersonic motion of the pulsar.

The luminosity, $L_{\rm comp}\sim 3\times 10^{32} d_4^2$ ergs s$^{-1}$
in the 0.5--8 keV band\footnote{We should note that our estimate for the
compact PWN luminosity is a factor of 3 lower than that obtained
by G03 for the ``core'' component. The discrepancy is possibly
caused by the much larger area from wich G03 extracted the spectrum
(1018 vs.\ 225 arcsec$^2$), which included some contribution from
the extended (``diffuse'') PWN, the pulsar, and field sources, such as Source X
in our Figures 1 and 2. Also, G03 extracted the background spectrum
from a region outside the extended PWN, while we used an annulus
filled with the extended PWN emission (see \S2.3).
 },
 the X-ray efficiency,
$\eta_{\rm comp}\equiv L_{\rm comp}/\dot{E}
\approx 1.2\times 10^{-4} d_4^2$,
  and the spectral slope, $\Gamma_{\rm comp}\approx
1.3$ (at the assumed $n_{\rm H,22}=1.0$) of the compact PWN
are not unusual for Vela-like PWNe
(e.g., Kargaltsev et al.\ 2007; Kargaltsev \& Pavlov 2007).
Also, similar to other Vela-like PWNe, the PWN luminosity
exceeds by a factor of a few the pulsar luminosity in the
same X-ray energy range,  $L_{\rm comp}\sim 4 L_{\rm PSR}$.

\subsubsection{Extended PWN}
The morphology of the extended PWN is even more puzzling than that
of the compact one. In the conventional picture of a bow-shock PWN,
we would expect to see an extended component
{\em behind} the compact PWN of
the moving pulsar,
which would look like a ``tail''. In B1823, however,
the PWN extends southward from the pulsar, almost
perpendicular to the proper motion direction,
similar to the extended diffuse PWN of the Vela pulsar
(Pavlov et al.\ 2003). This means that something
is deflecting (or has deflected) the shocked PW in this direction.
One could imagine a local north wind in the turbulent ambient medium,
but such a wind would have to blow with a
velocity of several hundred
km s$^{-1}$ (comparable to the pulsar velocity) to displace the PWN bubble.
If such a wind were blowing now, we would see the compact PWN elongated
southward from the pulsar, and, more importantly,
it would be not a usual wind but a shock, since the required velocity
almost certainly exceeds the sound speed. Thus, it seems we have
to assume that the PWN bubble was crushed/displaced by a passing
shock some time ago. A natural candidate for such a shock is the
reverse shock from the surrounding SNR that has propagated back
inward and collided with the PWN, displacing the PWN bubble from
the pulsar, as has been suggested by Blondin et al.\ (2001) for the
strongly dispaced Vela X component of the Vela PWN.
Such an explanation for the
one-sided extended PWN of B1823 has been proposed by G03, and we
currently do not see an alternative explanation. If this hypothesis
is correct, we may expect the extended PWN to consist of two components:
a strongly displaced one (similar to Vela X),
which represents a relic population of
relativistic electrons that have lost a substantial fraction of
their energy to synchrotron, inverse Comption, and adiabatic cooling,
and a younger one, which represents electrons that
entered the PWN after the inverse
shock passage.

Although the crushed/dragged PWN hypothesis looks attractive
and provides a plausible explanation for the extended X-ray PWN,
we should mention that it does not so readily expalin the huge,
about 70 pc at $d=4$ kpc, size of the
TeV PWN. Traveling 70 pc in time $t$ requires the average
velocity $v\approx 6,900\, (t/10\,{\rm kyr})^{-1}$ km s$^{-1}$, too high
for a reverse shock. Even if we assume $t=21$ kyr
(i.e., equal to the pulsar's spindown age), we still need
$v\approx 3,300$ km s$^{-1}$. Therefore, it is hard to imagine
how the reverse shock could have dragged the relic PWN so far,
even if it passed the pulsar very long time ago.
Perhaps, one should explore
another possibility, that during a substantial fraction of the pulsar's
lifetime the PWN electrons have been
transported southward from the pulsar
along an anisotropic large-scale
magnetic field.
Alternatively, one might assume a strongly anisotropic supply
of electrons into the PWN, e.g., through a one-sided jet
(which, however, is not seen in the images available).
At this point, however, these possibilities remain highly
speculative.

As has been noticed by G03, the extended PWN spectrum is softer
than the compact PWN spectrum, which is also confirmed by our
results (see Fig.\ 5). The difference of the photon indices,
$\Gamma_{\rm ext} - \Gamma_{\rm comp} \approx 0.5$, supports the hypothesis
that the extended X-ray PWN electrons have lost their energy to synchrotron
cooling after traveling a distance of a few parsecs.
The slope $\Gamma_{\rm ext}\approx
1.9\pm 0.1$ we found for the X-ray spectrum is very close to
the slope of the VHE $\gamma$-ray spectrum within $\approx 6'$ from the pulsar,
$\Gamma_{\gamma}=1.8$--2.0 (A06),
which confirms that both
the X-ray and $\gamma$-ray emission are due to the same electron population
with a power-law energy spectrum $dN_e/dE_e\propto E_e^{-p}$,
with $p=2\Gamma_{\rm ext}-1\approx 2.5$--3.
Interestingly, the TeV spectrum
at larger distances from the pulsar, where we do not see the
X-ray PWN, becomes even softer ($\Gamma_\gamma$ grows up to
$\approx 2.6$ at $60'$ from the pulsar).

Because we can see only part of the extended PWN in our ACIS
image (see \S2.2),
we can measure only a fraction of its X-ray luminosity
($L_{\rm ext}\sim 1\times 10^{33} d_4^2$ ergs s$^{-1}$ in the
3.4 arcmin$^2$ region shown in Fig.\ 1, {\em bottom right}).
The total extended PWN luminosity should be a factor of a few
higher. For instance, G03 found $L_{\rm ext} \sim 3\times 10^{33} d_4^2$
ergs s$^{-1}$ in a 33.4 arcmin$^2$ area, which corresponds to
the extended PWN efficiency $\eta_{\rm ext} \sim 10^{-3} d_4^2$,
a factor of $\sim 10$ higher than that of the compact PWN.

\subsection{Pulsar}
Although we have resolved the pulsar from the PWN, the small
number of pulsar counts detected and the strong
and not well determined  background
from the inner component of the compact PWN preclude a precise
spectral analysis. We can only estimate the pulsar
luminosity, $L_{\rm PSR} \sim 1\times 10^{32} d_4^2$ ergs s$^{-1}$
in the 0.5--8 keV band, which is a small fraction,
$\sim 3\times 10^{-5}$, of the pulsar's spindown power $\dot{E}$.
The slope of the pulsar spectrum, $\Gamma_{\rm PSR}\sim2$--3,
is strongly correlated with the ISM column density (see Fig.\ 7). It is
likely that the spectrum consists of two components, magnetospheric
and thermal, but the photon statistics does not allow one to separate
these components unambiguously. For $n_{\rm H,22}=1.3$ (at the
higher end of the plausible $n_{\rm H}$ range) and projected emitting areas
in the range of $(20$--$500)d_4^2$ km$^2$, we found $kT\sim
90$--150 eV and $L_{\rm bol} \sim (3$--$20)\times 10^{32} d_4^2$ ergs s$^{-1}$
for a possible BB component (higher luminosities correspond to lower
temperates and larger emitting areas).
These temperatures and luminosities are generally similar
to those of other Vela-like pulsars
(e.g., $kT\approx 130$ eV, $\mathcal{A}\approx 20 (d/300\,{\rm pc})^2$ km$^2$,
 $L_{\rm bol}\sim 2\times 10^{32}
(d/300\,{\rm pc})^2$ ergs s$^{-1}$ for the Vela pulsar; Pavlov et al.\ 2001).
However, because of the large uncertainties and  poorly constrained
$n_{\rm H}$, we cannot be sure even in the
detection of the thermal component,
not to mention measuring its parameters. Therefore, we should treat
the estimated bolometric luminosities of B1823 as upper limits
rather than actually measured values.
The most conservative of the above limits,
$L_{\rm bol}\sim 2\times 10^{33} d_4^2$ ergs s$^{-1}$,
corresponding to
$\mathcal{A}=500 d_4^2$ km$^2$ and $T\approx 100$ eV,
is close to the bolometric luminosities predicted by the so-called
``standard'' NS cooling model for the NS age of 20 kyr
(see, e.g., Fig.\ 2 in Tsuruta 1998).
Overall, we can conclude that B1823 looks like a typical Vela-like
pulsar, but a deeper high-resolution observation is needed to infer its
properties in more detail.

\subsection{The lack of a host SNR?}

We see no emission in our ACIS images that could be ascribed
to an SNR associated with B1823, and no such emission has been
reported from the previous X-ray (F96; G03) or radio
(Braun et al.\ 1989; Brogan et al.\ 2006) observations.
Using our proper motion measurement, we extrapolated
the pulsar trajectory back in time to check whether or not
it crosses a known SNR. The nearest SNR to the pulsar trajectory
is the $8'\times 8'$ shell SNR G18.16--0.16
 (Brogan et al.\ 2006) at a distance of $35'$ from the current
pulsar position.
Since
the extrapolated trajectory
misses the SNR center by $19'\pm 4'$,
we conclude that G18.16--0.16 is not related to B1823.

The failure to detect the host SNR in X-rays can be attributed
to the strong ISM absorption of the soft thermal emission from the
SNR interiors (see Kargaltsev et al.\ 2007, where a similar situation
is discussed regarding the Vela-like PSR B1800--21, which also lacks
an associated SNR).
Therefore, it is possible that B1823 is still within the interiors
of an evolved SNR, which is undetectable in X-ray or radio, but whose
pressure can contribute to the PWN confinement.

\subsection{Connection between B1823 and HESS\,J1825}

The important discovery of the VHE $\gamma$-ray spectral softening
with increasing distance from B1823
(A06) has provided further evidence that
this pulsar/PWN is the main source of the
ultrarelativistic electrons powering the TeV plerion.
Qualitative models for this VHE $\gamma$-ray
 source have been discussed in a number
of works
(A06; de Jager et al.\ 2005; de Jager 2007).
Generally, they interpret the $\gamma$-ray
 emission as produced by the Compton
 upscattering
of the CMB photons by electrons with energies $E_e \sim 25
(E_\gamma/1\,{\rm TeV})^{1/2}\,\, {\rm TeV} \sim 10$--100 TeV
[electron Lorentz factor $\gamma_e\sim (2$--$20)\times 10^7$],
supplied by the B1823 pulsar. The fact that the pulsar is located at the
northern boundary of the TeV plerion is explained by the effect of the
SNR reverse shock that has dragged the relic PWN southward.
(An alternative explanation,
that the TeV plerion represents electrons left behind the pulsar
moving northward, is invalidated by our measurement of the proper
motion.) The softening of the TeV spectrum with the distance from
the pulsar can be ascribed to the radiative losses (mostly to the
IC scattering losses) or some energy-dependent ``convection'', or  variation
of the particle injection spectrum with the pulsar's age.
As A06 note, even at very low ambient magnetic fields, $B\lesssim 3\, \mu$G,
when the IC energy losses dominate the synchrotron losses, the
 $\gamma$-ray
efficiency, $\eta_\gamma =L_\gamma/\dot{E}\approx 0.1$,
is unrealistically high if we use
the current value of $\dot{E}$.
This supports the hypothesis that the
VHE $\gamma$-ray emission is produced by relic electrons
emitted by the pulsar when it was very young and much more
powerful.

Modelling of the TeV plerion is beyond the scope of
this work. We only note some open problems with the current
interpretation of this remarkable object. First, as we have mentioned
in \S4.1, it is difficult to reconcile the very large size of the
TeV plerion with the hypothesis that the relic PWN was dragged by
the SNR reverse shock to its current position. In principle,
other sources of ultrarelativistic electrons
within the  $1^\circ$ size field could contribute to the observed
TeV emission and increase the apparent size of the TeV source.
Our analysis of several X-ray sources in the brightest
part of the TeV plerion has not shown any obvious candidates
(see \S2.5), but one cannot rule out the possibility that there
are such sources outside the small ACIS field-of-view.

Another possible problem with the current interpretation of
the TeV source is the spatial distribution of its brightness and spectrum.
The brightest part of the TeV plerion is offset by $\sim 10'$
($\sim 12$ pc in the plane of the sky)
south-southwest of the pulsar (see Fig.\ 1 of A06),
while
the plerion extends $\sim 50'$ further in the same direcion
(almost perpendicular to the direction of the pulsar's proper motion).
One could speculate
that the relic PWN was displaced by $\sim 10'$
some time ago [which would require a reverse shock speed of $\sim
1000 (t/10\,{\rm kyr})^{-1}$ km s$^{-1}$], and its relativistic electrons
have been diffusing from that position. It is however, unclear as to
why they diffused not radially but preferentially in the
south-southwest direction.
If the passing shock not just displaced
the relic PWN to a certain distance but smeared it all over the
whole extent of the currently seen TeV plerion, then a very high
shock speed would be required (see \S4.1).
Furthermore, the spectral map of HESS\,J1825 (see Figs.\ 5 and 6 of
A06) shows that the VHE $\gamma$-ray spectrum softens radially not from the
maximum of the brightness distribution but from the pulsar position.
This hints that at least some contribution to the TeV photons
comes from the same young electrons that are responsible for the
extended X-ray PWN of a much smaller size. This conjecture is
further supported by the fact that the spectral slopes of the
X-ray (synchrotron) and $\gamma$-ray (IC) spectra are
about the same at a few arcminute distance from the pulsar, which
suggests that they are generated by the same population of
electrons. On the other hand, if the TeV $\gamma$-ray emission
is due to the IC scattering of the CMB photons, and the relativistic
electrons are mostly supplied by the pulsar, one should expect the
TeV brightness to peak at the pulsar position, not $10'$ from the pulsar.
This inconsistency might be alleviated assuming that
in the immediate vicinity of the pulsar,
where the electrons are more energetic, they upscatter the CMB
photons in the Klein-Nishina regime, which results in a lower
brightness of the IC scattered radiation, but it remains to be
seen from detailed models whether or not this assumption is realistic.
Overall, it seems plausible at this point that, at least in the
extended X-ray PWN region, the TeV emission can be produced by two
electron populations of different ages. To infer the properties
of these populations, deep observations with {\sl XMM-Newton}
and
{\sl GLAST} of a region from which {\em both} X-ray
and VHE $\gamma$-ray emission is seen would be very helpful.

Another issue related to the nature of the TeV plerion and its
connection to the B1823 pulsar/PWN is whether the TeV photons
are produced by upscattering of the CMB or some other radiation.
Close to the Galactic plane, where HESS\,J1825 is situated,
a factor of a few higher radiation energy density can be provided
by IR emission from interstellar dust and Galactic starlight.
For instance,
at the galactocentric distance
of 4 kpc, the models of interstellar radiation by Strong
et al.\ (2000) give the energy densities of 0.6 and 2.7 eV cm$^{-3}$ for
these two components, respectively,
versus 0.26 eV cm$^{-3}$ for
the CMB. Moreover, if there are very luminous sources of radiation
(such as
molecular/dust complexes
in star-forming regions) in the vicinity of the source of
ultrarelativistic electrons, the high radiation energy density
around these sources can strongly enhance the TeV luminosity
(see, e.g., Kargaltsev \& Pavlov 2007).
In this regard, the bright radio-IR Source A (see Fig.\ 3),
whose spectrum peaks at $\lambda \gtrsim 100\,\mu$m, might be
of some interest.
With the
bolometric flux $F_A \sim 2\times 10^{-8}$ ergs cm$^{-2}$ s$^{-1}$,
its radiation energy density exceeds the ambient radiation energy
density $U_{\rm amb}$ at angular distances $\theta < (F_A/cU_{\rm
abm})^{1/2} \sim 2.2' (U_{\rm amb}/1\,{\rm eV\, cm}^{-3})^{-1/2}$. If
the distance to Source A is about the same as to B1823, and the TeV
emission is produced by the IC scattering of IR photons with energies
around 0.03 eV (which would require electron energies $E_e \sim
1$--20 TeV), then Source A could create a few arcminute size
brightening. Moreover, if there are hot and massive stars hidden
within this dust/molecular complex of $\sim 4 d_4$ pc size, their
shocked winds could be additional sources of TeV electrons, further
enhancing the VHE $\gamma$-ray emission from this region. (VHE
$\gamma$-rays apparently produced by stellar winds in the open
stellar cluster Westerlund 2 have recently been detected by
Aharonian et al.\ 2007.)
  Although this hypothesis looks rather
speculative at this point, and Source A can not be responsible for the
bulk of the detected TeV radiation, dedicated observations of this
region in the IR and radio would help evaluate the Source A contribution
to HESS\,J1825.

\section{Conclusions}

The high-resolution {\sl Chandra} ACIS observation has allowed us to
resolve the B1823 pulsar from its PWN, image the PWN structure, and measure
the spectra of the pulsar and the PWN components. We found that the
pulsar's X-ray luminosity,
$L_{\rm PSR}\sim 0.8\times 10^{32}$ ergs s$^{-1}$ in the 0.5--8 keV band,
and spectrum
 are similar to those of other Vela-like
pulsars, but the small number of pulsar counts and the contamination
from the inner PWN component do not allow an unambiguous separation
of the thermal and nonthermal components of the pulsar radiation.

We have measured the pulsar's proper motion and found its transverse
velocity, $v_\perp = 443\pm 48$ km s$^{-1}$ at $d=4$ kpc, close to
an average velocity of young radio pulsars. We found no indication
of an SNR, in neither X-rays nor radio, between the current
position of the pulsar and its birthplace, $\sim 8'$ westward.

We have confirmed
the presence of the compact and extended PWNe of different shapes,
surface brightnesses, and spectra.
The compact PWN, of $\sim 0.5\times 0.2$ pc$^2$ size in the plane
of the sky, is elongated
approximately along the proper motion direcion, with the pulsar
close to its leading edge. This suggests that the postshock pulsar
wind in the compact PWN, responsible for its X-ray emission, is
confined by the ram pressure due to the supersonic motion of the pulsar.
The X-ray spectrum and luminosity,
$L_{\rm comp}\sim 3\times 10^{32}$
ergs s$^{-1}$, of the compact PWN are similar to those of other
Vela-like PWNe.

The X-ray surface brightness distribution within the compact PWN
is very nonuniform. In particular, we see a much brighter inner
PWN component, $\sim 0.14\times 0.06$ pc$^2$, around the pulsar,
inclined by $\sim 50^\circ$ with respect to the direction of the proper
motion. Such a structure indicates that the pulsar wind is intrincically
anisotropic. If the inner PWN component represents the shocked
equatorial outflow
from the pulsar magnetosphere, the pulsar's spin axis is strongly
inclined to the proper motion direction, in contrast to many other
pulsars.

The extended PWN is mostly seen south-southwest of the pulsar
(up to $2.4'$ from the pulsar, in the ACIS image), almost
perpendicular to the proper motion direction. Its softer spectrum
indicates that its X-ray emission is generated by older electrons
that have lost part of their energy because of the radiative cooling.
The southward displacement of the extended X-ray PWN with respect to the
pulsar might be
caused by a strong pressure gradient in the
ambient medium and/or by
the reverse SNR shock, which
has swept the PWN material southward.

The extended X-ray PWN is apparently connected to the
VHE $\gamma$-ray PWN that extends in the same direction but to
a much larger distance from the pulsar, up to 70 pc in the plane
of the sky. The $\gamma$-ray emission is likely due to the IC
scattering of background radiation off the relativistic electrons
with energies up to 100 TeV supplied by the B1823 pulsar.
However, the
very high
luminosity, $L_\gamma \sim 3\times
10^{35}$ ergs s$^{-1}$, can only be explained if the
bulk of $\gamma$-radiation is generated by relic electrons,
produced long ago, when the pulsar was much more powerful.
The relic electrons could be displaced from the pulsar by a reverse
SNR shock, but
it remains unclear as to why the $\gamma$-ray PWN is so large.
We have not found X-ray sources in the brightest part of the $\gamma$-ray
PWN which could significantly contribute to the $\gamma$-radiation.
Although the IR-radio images of the field show a star-forming region
in the brightest part of the $\gamma$-ray PWN, which could enhance
the $\gamma$-ray emission by supplying additional photons for the
IC scattering, and perhaps even additional relativistic electrons,
it can be responsible only for a small fraction of the observed
VHE $\gamma$-radiation.
To better understand the multiwavelength nature of the complex
B1823 PWN, deep X-ray and IR observations of the field would
be particularly useful.

\acknowledgements
Our thanks are due to Sachiko Tsuruta, PI of the {\sl Chandra} observation,
Marcus Teter, who performed the initial analysis of
the X-ray data, and Vlad Kondratiev, who inspected the archival radio data.
Support for this work was provided by the National Aeronautics and
Space Administration through Chandra Award Number AR5-606X
 issued by the Chandra X-ray Observatory Center,
which is operated by the Smithsonian Astrophysical Observatory for and
on behalf of the National Aeronautics Space Administration under
contract NAS8-03060.
This work was also partially supported by NASA grant NAG5-10865.

\clearpage

\begin{table}[]
\caption[]{PL fits to the PWN spectrum} \vspace{-0.5cm}
\begin{center}
\begin{tabular}{ccccccc}
\tableline\tableline Region &
$\mathcal{N}$\tablenotemark{a}  &
$\Gamma$ & $F$\tablenotemark{b} & $F_{\rm unabs}$\tablenotemark{c} & $\chi^2$/do
f  \\
\tableline
  Compact PWN          &
$2.14\pm 0.25$      &
$1.33^{-0.13}_{+0.11}$ & $1.1\pm0.1$ & $1.7\pm 0.1$ &  0.93/18 \\
 Extended PWN      &
$9.0\pm 1.0$ &  $1.87\pm0.14$ & $2.5\pm 0.2$ & $4.5^{+0.3}_{-0.2}$ & 1.28/13 \\
      \tableline
\end{tabular}
\end{center}
\tablecomments{The fits are for fixed $n_{\rm H,22}\equiv n_{\rm
H}/(10^{22}\,{\rm cm}^{-2}) =1.0$. The uncertainties are given at 68\%
confidence level for a single interesting parameter.}
\tablenotetext{a}{Spectral flux in units of $10^{-5}$ photons
cm$^{-2}$ s$^{-1}$ keV$^{-1}$ at 1 keV.}
\tablenotetext{b}{Observed energy flux in units of $10^{-13}$ ergs cm$^{-2}$ s$^
{-1}$, in the 0.7--7 keV band.}
\tablenotetext{c}{Unabsorbed energy flux in units of $10^{-13}$ ergs cm$^{-2}$ s
$^{-1}$, in the 0.5--8 keV band.}
\end{table}

\begin{table}[]
\caption[]{Fits to the pulsar spectrum} \vspace{-0.5cm}
\begin{center}
\begin{tabular}{ccccccc}
\tableline\tableline Model & $n_{\rm H,22}$   &
$\mathcal{N}$\tablenotemark{a} or $\mathcal{A}$\tablenotemark{b} &
$\Gamma$ or $kT$\tablenotemark{c} & $C$  & $L_{\rm X}$ or
$L_{\rm bol}$\tablenotemark{d} \\
\tableline
PL & $0.7$ &
$6.8_{-1.6}^{+2.1}$ & $1.98_{-0.36}^{+0.40}$ & $639$ &
0.6 \\
PL & $1.0$ & $11.5_{-3.2}^{+4.2}$ & $2.43_{-0.40}^{+0.53}$   &  $646$    &
0.8 \\
PL & $1.3$ & $19.6^{+8.0}_{-6.2}$ & $2.92_{-0.54}^{+0.62}$   &  $657$    &
1.1 \\
PL+BB(PL)    &     $1.3$ & $7.5^{+3.0}_{-1.9}$ & $1.9\pm 0.6$  &  $650$   &
0.7 \\
 PL+BB(BB)  &  $1.3$ & $500$ &
$97^{+4}_{-5}$   &  $650$   & $18$   \\
PL+BB(PL)    & 1.3 &    $5.0^{+5.1}_{-3.1}$ & $1.7\pm 0.7$ & 651 & 0.6 \\
 PL+BB(BB)  &  $1.3$ & $20$ & $139^{+9}_{-6}$ & 651 & $3.0$  \\
  \tableline
\end{tabular}
\end{center}
\tablecomments{The fits are for fixed $n_{\rm H,22}\equiv n_{\rm
H}/10^{22}$ cm$^{-2}$. The uncertainties are given at 68\%
confidence level for a single interesting parameter. In each case the
fits were done using the C-statistics (Cash 1979) and the unbinned
source and background spectra with the total of 1052 channels.
 }
 \tablenotetext{a}{Spectral flux in units of $10^{-6}$
photons cm$^{-2}$ s$^{-1}$ keV$^{-1}$ at 1 keV.}
\tablenotetext{b}{Projected area of the emitting region for the BB
model, in units of km$^2$, fixed in the fits.}
\tablenotetext{c}{BB temperature in eV.}
\tablenotetext{d}{Unabsorbed PL luminosity in the 0.5--8 keV band or
bolometric BB luminosity, in units of $10^{32}$ ergs s$^{-1}$.}
\end{table}

\begin{table}[]
\caption[]{Positions and spectral properties of the field X-ray sources}
\vspace{-0.5cm}
\begin{center}
\begin{tabular}{cccccccccc}
\tableline\tableline Source\tablenotemark{a} & R.A. & Decl. &
$\delta r$\tablenotemark{b} & $n_{\rm H,22}$   &
$\mathcal{N}$\tablenotemark{c}  &
$\Gamma$
& $F_{\rm obs}$\tablenotemark{d}  & $F_{\rm unabs}$\tablenotemark{e} \\
\tableline
X & 18:26:14.43 & $-$13:34:48.0 &
$0.4''$ & $2.9^{+2.9}_{-1.6}$ & $0.10_{-0.05}^{+0.08}$ & $0.2\pm0.5$              &
$2.9\pm0.5$ & $3.7$ \\
s1 & 18:26:14.06 & $-$13:40:35.8 &
$0.5''$ & $4.3^{+1.1}_{-0.8}$ & $2.2_{-1.0}^{+1.7}$ & $2.1\pm0.5$
& $3.2\pm0.4$ & $11.9$ \\
s2 & 18:26:17.18 & $-$13:41:12.7 &
$0.5''$ & $0.9^{+0.7}_{-0.3}$ & $1.8_{-0.4}^{+0.6}$ & $3.1^{-0.3}_{+0.5}$
 & $1.5\pm0.3$  & $7.0$ \\
s3 & 18:26:17.99 & $-$13:42:27.5 &
$0.5''$ & $0.5^{+0.8}_{-0.4}$ & $0.5\pm0.2$   & $1.9\pm0.4$   &
$1.7\pm0.4$  & $2.7$ \\
s4 & 18:26:20.88 & $-$13:44:25.7 &
$0.8''$ & $1.2^{+0.6}_{-0.4}$ & $3.5_{-0.6}^{+0.7}$ & $3.1\pm0.3$
  & $2.8\pm0.3$  & $14.2$ \\
s5 & 18:26:24.76 & $-$13:37:06.2 &
$0.5''$ & $3.0^{+2.6}_{-2.3}$ & $0.8_{-0.4}^{+0.6}$ & $1.8\pm0.5$
& $1.9\pm0.3$ & $4.6$ \\
i1 & 18:25:40.55 & $-$13:40:11.2 &
$0.6''$ & $1.1^{+1.2}_{-1.0}$ & $1.3_{-0.3}^{+0.4}$ & $1.9\pm0.3$
  & $3.2\pm0.6$ & $6.4$ \\
 \tableline
\end{tabular}
\end{center}
\tablecomments{ The uncertainties of the fitting parameters
 are given at 68\% confidence level
for a single interesting parameter.
 }
\tablenotetext{a}{Source designation as shown in
Figs.\ 1 and 8.}
\tablenotetext{b}{Radius of the position error circle at the 68\% confidence level.}
\tablenotetext{c}{Spectral flux in units of $10^{-5}$
photons cm$^{-2}$ s$^{-1}$ keV$^{-1}$ at 1 keV.}
\tablenotetext{d}{Observed flux in the 1--8 keV band, in units of
$10^{-14}$ ergs  cm$^{-2}$ s$^{-1}$.}
\tablenotetext{e}{Best-fit value of the unabsorbed
PL flux in the 0.5--8 keV band, in units of $10^{-14}$ ergs  cm$^{-2}$
s$^{-1}$.}
\end{table}

\end{document}